\documentclass[aps,preprint]{revtex4}
\usepackage{amsmath}
\usepackage{amssymb}
\usepackage{mathrsfs}
\usepackage{xcolor}
\usepackage{graphicx}
\usepackage{subfigure}
\usepackage{enumerate}
\usepackage{float}

\begin{document}
	\title{Thermodynamics of the RN-AdS black hole with cloud of strings and quintessence in stationary and free-fall frame in rainbow gravity}
	

	\author{Siyuan Hui$^{a}$}
	\email{huisiyuan@stu.scu.edu.cn}
	\author{Benrong Mu$^{a,b}$}
	\email{benrongmu@cdutcm.edu.cn}
	\author{Yuzhou Tao$^{a}$}
	\email{taoyuzhou@stu.scu.edu.cn}
	\author{Jun Tao$^{a}$}
	\email{taojun@scu.edu.cn}
	
	\affiliation{$^{a}$Center for Theoretical Physics, College of Physics, Sichuan University, Chengdu, 610064, PR China\\
	$^{b}$Physics Teaching and Research section, College of Medical Technology, Chengdu University of Traditional Chinese Medicine, Chengdu, 611137, PR China}

	\begin{abstract}
	
	In this paper, we investigate the thermodynamic properties of the RN-AdS black hole with cloud of strings and quintessence in rainbow gravity with the stationary (ST) orthonormal frame and the free-fall (FF) orthonormal frame. After the SF and the FF rainbow metric is obtained, we get the Hawking temperature and the entropy, and their physical meanings are discussed. We find that, for the ST rainbow RN-AdS black hole with cloud of strings and quintessence, the effect of rainbow gravity is to increase the Hawking temperature but decrease the entropy of the black hole. However, for the FF rainbow case, rainbow gravity turns out to decrease the Hawking temperature but increase the entropy of the black hole, which seems that the effects rainbow gravity has are quite model-dependent.
	
	\end{abstract}
	\maketitle
	\tableofcontents
	
	\section{Introduction}

	In 1974, Stephen Hawking found that the Schwarzschild black hole emits radiation at a temperature just like an ordinary blackbody, which was later named as the Hawking radiation\cite{Hawking:1975vcx}. It was a very remarkable prediction of the quantum field theory in curved spacetime, which has contributed to studies about thermodynamics of black holes. Soon after this discovery, a possible trans-Planckian problem was realized by Unruh\cite{Unruh:1976db}. It shows that the Hawking radiation seems to start from the modes of huge initial frequencies, which are far beyond the Planck mass and experience exponential high gravitational red shifting near the horizon of black holes. Therefore, Stephen Hawking’s prediction depends on the validity of quantum field theory to arbitrary high energy in curved spacetime. And it gives out the question that whether any unknown nature of physics at the Planck scale could strongly affect the Hawking radiation.
	
	On the other hand, the mathematical model of general relativity is dependent on the smooth manifold, which breaks down when spacetime is measured at energies of the order of Planck energy\cite{Maggiore:1993rv, Park:2007az}. Most of the quantum gravity theories suggest that the standard energy-momentum dispersion relation would get modified under the UV limit\cite{tHooft:1996ziz, Kostelecky:1988zi, Amelino-Camelia:1997ieq, Gambini:1998it, Carroll:2001ws, Horava:2009uw, Horava:2009if}, which have been constructed with the light speed and the Planck energy. This contributes to the modified energy-momentum dispersion relations (MDR). And this generalization of special relativity is called double special relativity (DSR) \cite{Kimpton:2013zb, Lopes:2015bra, Magueijo:2002am}.
	
	Modified dispersion relation in the rainbow gravity attracts great attention from the semi-classical viewpoint of the loop quantum gravity theory\cite{Gambini:1998it, Alfaro:2001rb, Sahlmann:2002qk, Smolin:2002sz, Smolin:2005cz}. Such modifications can be found in threshold anomaly of ultrahigh cosmic rays and TeV photons\cite{Amelino-Camelia:1997ieq, Amelino-Camelia:1997wnq, Colladay:1998fq, Coleman:1998ti, Amelino-Camelia:1999iec, Amelino-Camelia:2000bxx, Jacobson:2001tu, Jacobson:2003bn}. Meanwhile, threshold anomaly is not a general feature of the modified dispersion relationship, but they can only be predicted by modified dispersion relation scenarios with a preferred reference frame\cite{Amelino-Camelia:2002kwj}. In fact, the modified dispersion relations are dependent on the doubly special relativity\cite{Amelino-Camelia:2000cpa, Amelino-Camelia:2000stu, Amelino-Camelia:2003cem, Amelino-Camelia:2003xax, Magueijo:2001cr, Magueijo:2002am}.
	
	For this problem, it used to be claimed that the nonlinear Lorentz transformation in the momentum space was necessary to keep the double invariant constants. Later, DSR theory makes the Planck length a new invariant scale and gives out the nonlinear Lorentz transformations in momentum spacetime\cite{Amelino-Camelia:2000cpa, Amelino-Camelia:2000stu, Magueijo:2001cr, Magueijo:2002am}. 
	Specifically, the modified energy-momentum dispersion relation of particle energy $E$ and momentum $p$ can take the following form
	\begin{equation}
		E^2 F(E/E_p)^2-p^2 G(E/E_p)^2=m^2.
	\end{equation}
	where $E_{p}$ is the Planck energy.Amelino-Camelia\cite{Amelino-Camelia:1996bln,Amelino-Camelia:2008aez} proposed a popular choice of solution, which gives
	\begin{equation}
		\begin{aligned}
			F(x)=1,
			G(x)=\sqrt{1-\eta x^n}.
		\end{aligned}
	\end{equation}
	It is compatible with some results obtained in the loop quantum gravity method and reflects those obtained in $\kappa$-Minkowski spacetime and other noncommutative spacetimes. The phenomenological meaning of this "Amelino-Camelia (AC) dispersion relation" was also reviewed\cite{Amelino-Camelia:2008aez}. Subsequently, Magueijo and Smolin proposed that the spacetime background felt by a test particle depends on the test particle’s energy\cite{Magueijo:2002xx}. Thus, the energy of the test particle deforms the background geometry and eventually leads to modified dispersion relations. According to the modified dispersion relations under the generalized uncertainty principle, the second law of black hole thermodynamics is proved to be valid by modifying a relation between the mass and the temperature of the black hole\cite{Amelino-Camelia:2005zpp}. Moreover, Garattini gave out the result that the brick wall could be eliminated via appropriate rainbow functions\cite{Garattini:2009nq}. Additionally, the stationary(ST) orthonormal frame and the free-fall(FF) orthonormal frame have been proposed to expand the effects of rainbow gravity, which has been extensively studied to explore various aspects of black holes and cosmology\cite{Galan:2004st, Hackett:2005mb, Aloisio:2005qt, Amelino-Camelia:2013wha, Barrow:2013gia, Garattini:2011hy, Garattini:2011fs, Garattini:2014rwa, Mu:2019jjw, Gangopadhyay:2016rpl, Mu:2015qna, Gim:2015yxa, Ali:2014zea, Kim:2016qtp, Ling:2005bp, MahdavianYekta:2019dwf}.
	
	In this paper, we study the quantum gravity effect on the RN-AdS black hole with cloud of strings and quintessence in rainbow gravity with the stationary(ST) orthonormal frame and the free-fall(FF) orthonormal frame. The remainder of our article is summarized as follow. In Section \ref{222}, we review the metric of the RN-AdS black hole with cloud of strings and quintessence, and obtain the first law of its thermodynamics. In Section \ref{444}, the Hawking temperature and the entropy of the ST rainbow RN-AdS black hole with cloud of strings and quintessence are given. In Section \ref{666}, the 4-velocity of the black hole is obtained to get the Hawking temperature and the entropy of the FF rainbow RN-AdS black hole with cloud of strings and quintessence. In Section \ref{777}, the discussion and conclusion is given. Throughout the paper we take geometrized units $c = G = k_b = 1$, where the Planck constant $\hbar$ is square of the Planck mass $m_p$.

	\section{The RN-AdS black hole with cloud of strings and quintessence}\label{222}
	The static spherically symmetric metric of the RN-AdS black hole with cloud of strings and quintessence can be written as follow\cite{deMToledo:2018tjq, Toledo:2018hav, DiaseCosta:2018xyj, Toledo:2019amt, Sakti:2019iku,  Cai:2019nlo, Guo:2019hxa,  Ali:2019mxs, Chabab:2020ejk, Singh:2020tkf, Toledo:2020xnt, Liang:2020hjz,  Liang:2020uul, Ali:2021xrx, Yin:2021fsg, Younesizadeh:2021zho, Yin:2021akt,  Ali:2021fnz,  Cardenas:2021eri, He:2021aeo, Al-Badawi:2022uwh}
	\begin{equation}\label{1}
		ds^2 = -f(r) dt^2 +\frac{1}{f(r)} dr^2 + r^2 (d\theta^2 + \sin ^2 \theta d\phi^2),
	\end{equation}
	where
	\begin{equation}\label{2}
		f(r) = 1-a-\frac{2 M}{r} +\frac{Q^2}{r^2} -\frac{\alpha}{r^{3 \omega + 1}} -\frac{\Lambda r^2}{3}.
	\end{equation}
	Here $M$ and $Q$ represent the mass and the charge of the black hole. The Hawking temperature of the black hole becomes
	\begin{equation}\label{10}
		T = \frac{f'(r_+)}{4 \pi} = \frac{1}{4 \pi} (\frac{2 M}{{r_+}^2}-\frac{2 Q^2}{{r_+}^3}+\frac{(3 \omega + 1) \alpha}{r_+^{3 \omega + 2}}+\frac{2 r_+}{l^2}).
	\end{equation}
	where the cosmological constant $\Lambda$ is
	\begin{equation}
		\Lambda = -\frac{3}{l^2},
	\end{equation}
	and the electric potential is
	\begin{equation}
		\varphi =\frac{Q}{r_+}.
	\end{equation}
	Then the first law of thermodynamics of the charged RN-AdS black hole can be written as
	\begin{equation}\label{12}
		dM= TdS +\varphi dQ.
	\end{equation}

	As the angular momentum $L$ and energy $E$ of the particle are conserved, so we can apply separated variable method on the action and employ the following ansatz for the action $I$
	\begin{equation}
		I=-E t+W(r)+\Theta(x)+\zeta.
	\end{equation}
	Hamilton-Jacobi equation is given\cite{Mu:2019jjw}
	\begin{equation}\label{3}
			p_{\mu}p^{\mu}=g^{\mu \nu}\partial_{\mu}I\partial_{\nu}I=-m^2.
	\end{equation}
	Using the metric and equation(\ref{3}), we can get the Hamilton-Jacobi equation
	\begin{equation}\label{4}
		-\frac{1}{f(r)} E^2+ f(r) Pr^2+\frac{1}{r^2}h^{ab}(x) \partial_{a}\Theta(x) \partial_{b}\Theta(x)=-m^2,
	\end{equation}
	where
	\begin{equation}
		Pr \equiv \partial_{r}I=\partial_{r}W(r).
	\end{equation}
	Here we set
	\begin{equation}
		h_{ab}(x) dx^{a} dx^{b}=d\theta^2 + \sin ^2 \theta d\phi^2.
	\end{equation}
	The method of separation of variables gives the differential equation for $\Theta(x)$
	\begin{equation}
		h^{ab}(x) \partial_{a}\Theta(x) \partial_{b}\Theta(x)=\lambda,
	\end{equation}
	where $\lambda$ is a constant and determined by $h^{ab}(x)$. In this case, equation(\ref{4}) can be written as
	\begin{equation}
		-\frac{1}{f(r)} E^2+ f(r) Pr^2+\frac{\lambda}{r^2}=-m^2.
	\end{equation}
	From this equation we can solve $Pr$
	\begin{equation}
		Pr_{\pm}=\pm f(r)^{-1} [E^2-f(r) (\frac{\lambda}{r^2}+m^2)]^{\frac{1}{2}}.
	\end{equation}
	Using the residue theory for the semi-circle\cite{Mu:2019jjw}, we get
	\begin{equation}\label{8}
		ImI_{\pm}=\int Pr_{\pm} dr.
	\end{equation}
	The $\pm$ denotes outgoing/ingoing solutions, and $ImI_{\pm}$ means the imaginary part of action of probed particle. The $Pr$ is the momentum along the trajectory, which can be solved by Hamiton-Jacobi equation.

	\begin{figure}[htbp]
		\centering
		\subfigure[$a=0.025$, $\alpha=0.025$.]{
			\includegraphics[scale=0.61]{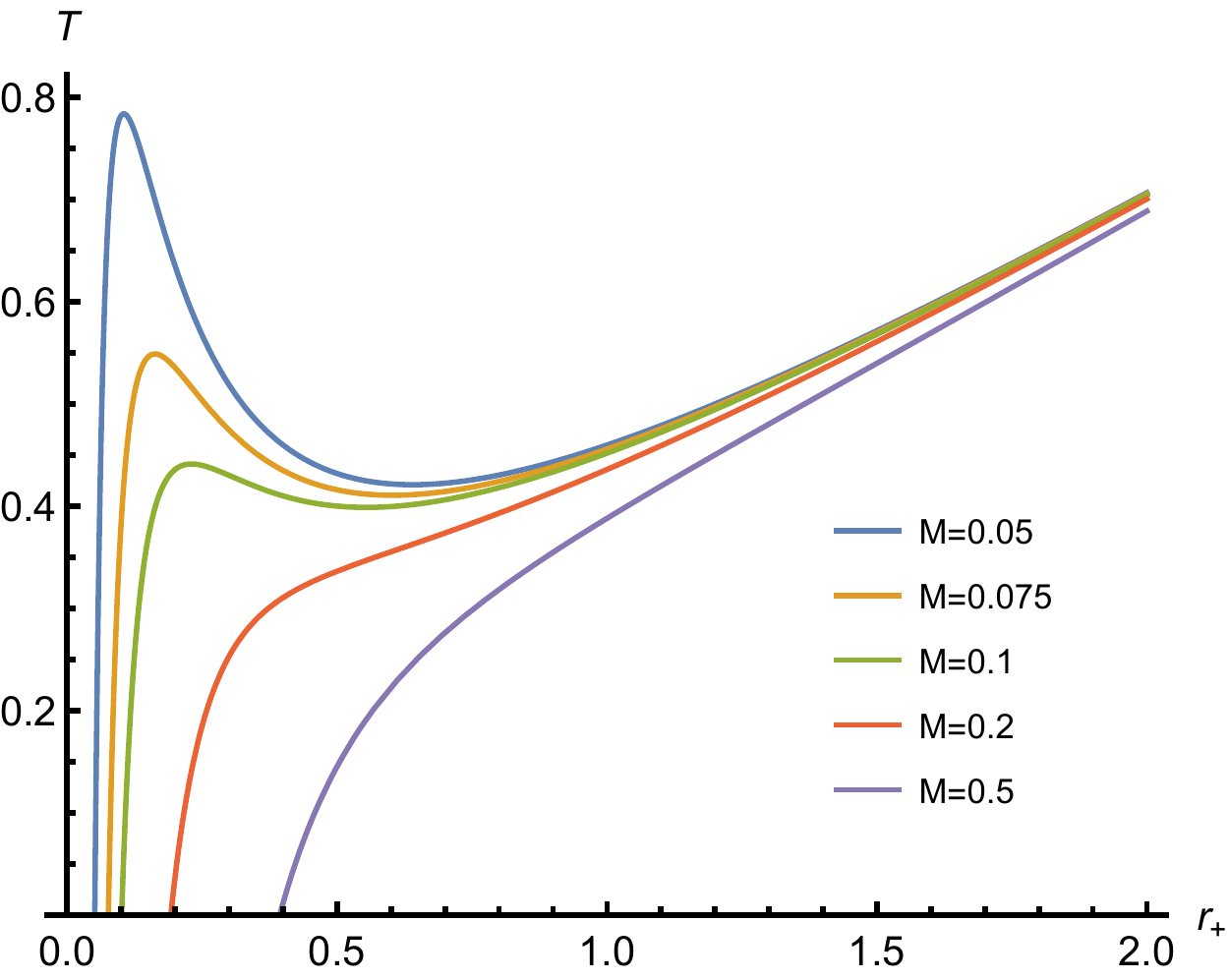}}
		\quad
		\subfigure[$M=0.05$, $\alpha=0.025$.]{
			\includegraphics[scale=0.61]{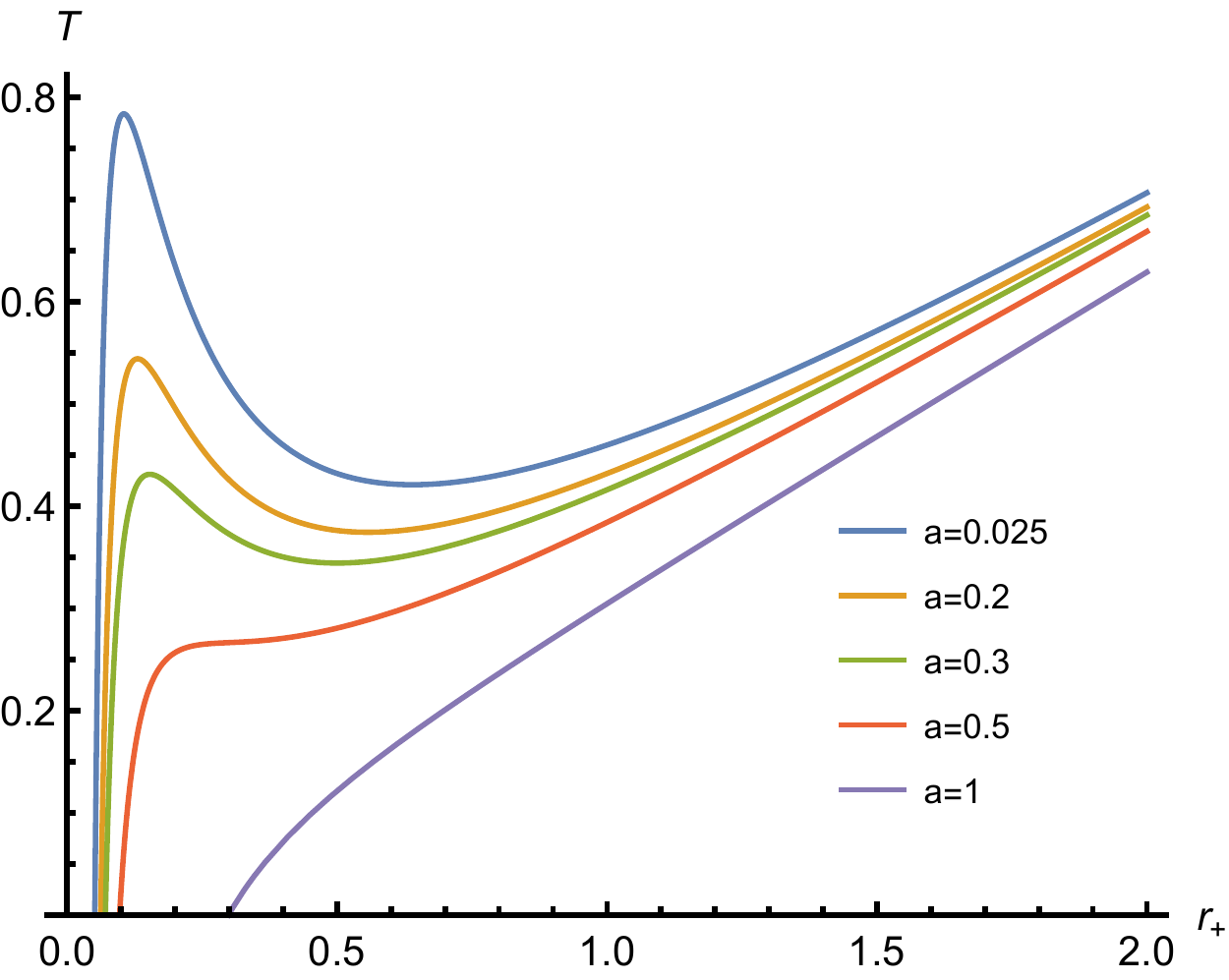}}
		\quad
		\subfigure[$M=0.05$, $a=0.025$.]{
			\includegraphics[scale=0.61]{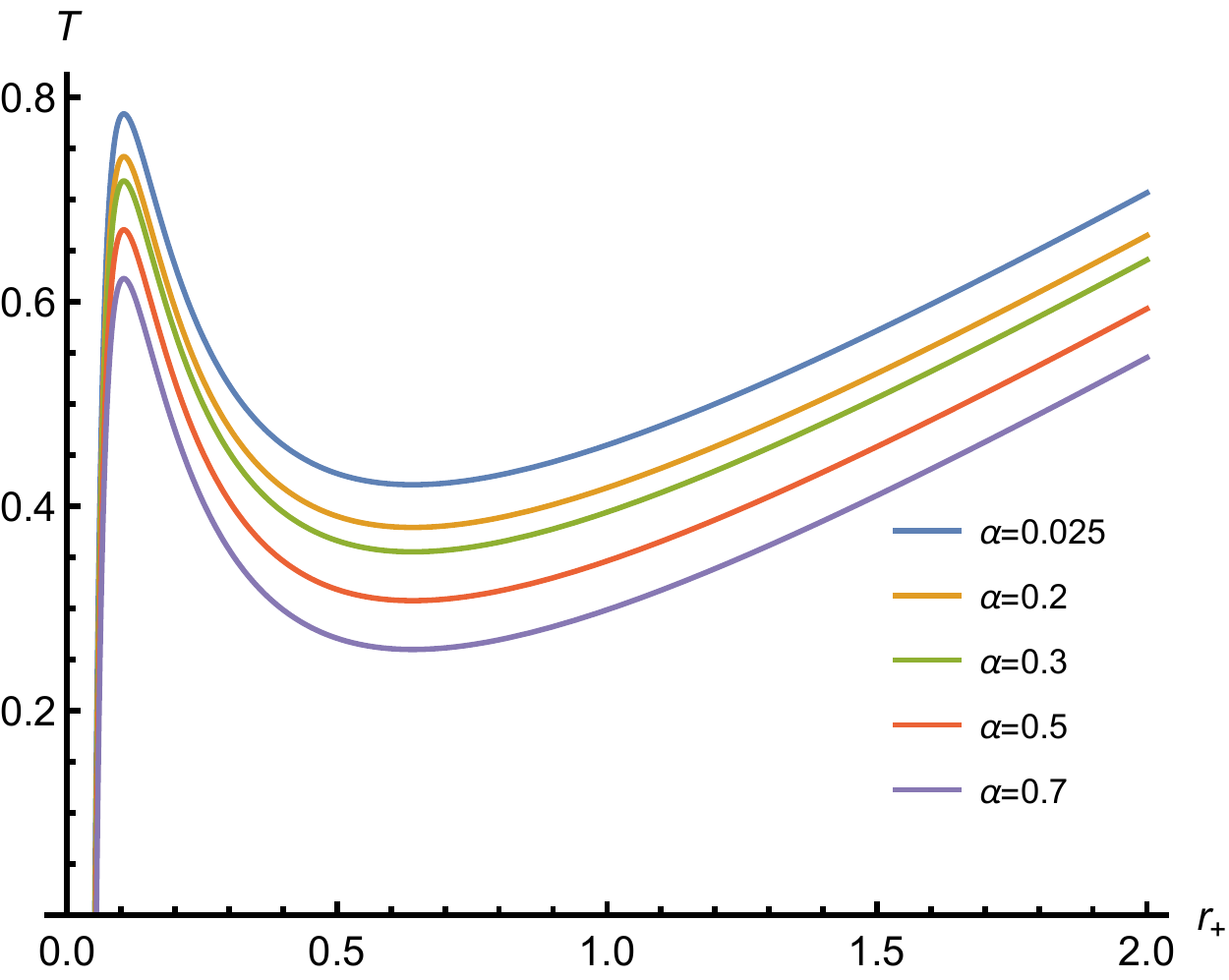}}
		\quad
		\subfigure[change $M$, $a$ and $\alpha$.]{
			\includegraphics[scale=0.61]{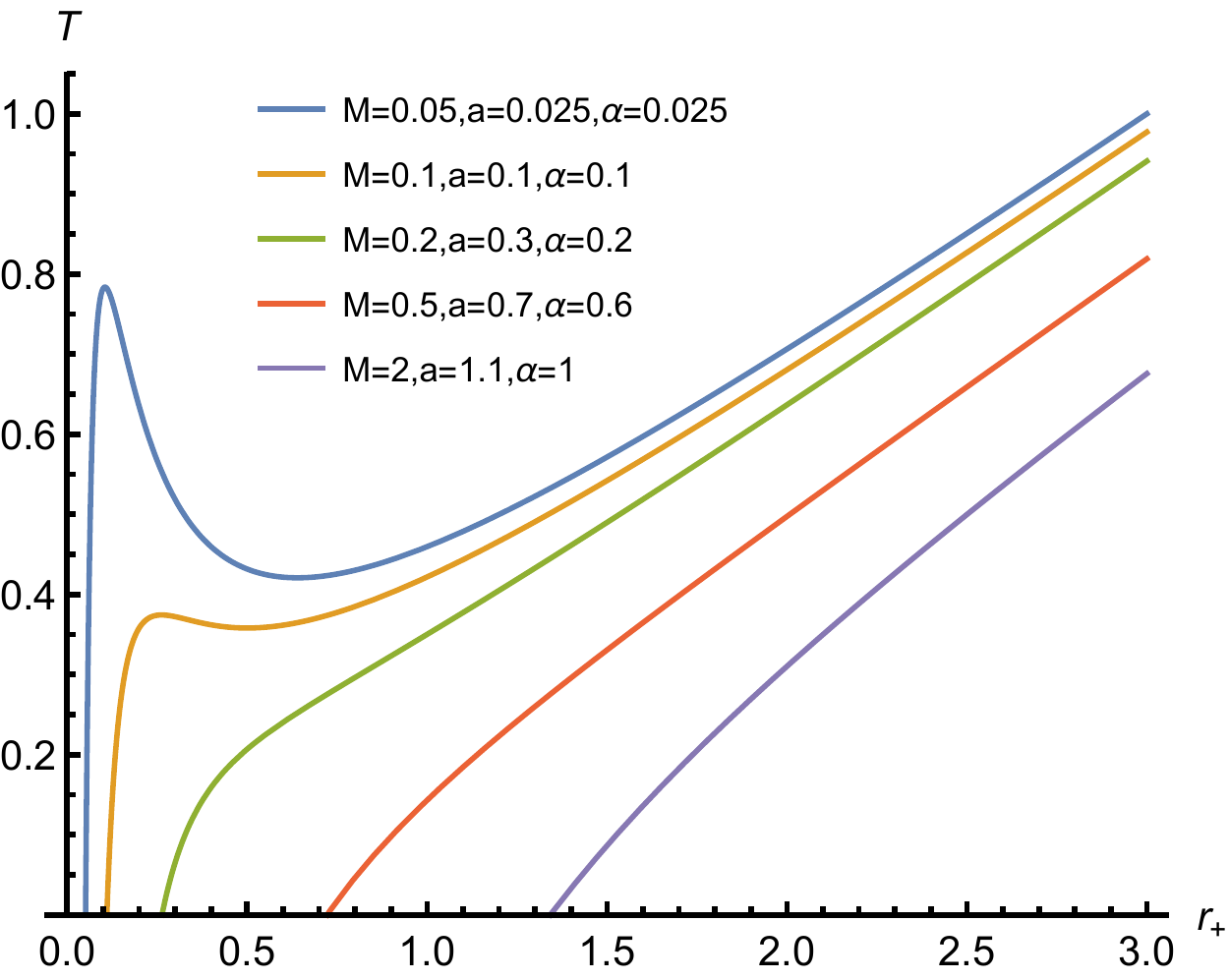}}
		\caption{Plots of the Hawking temperature $T(r_+)$ of a RN-AdS black hole with cloud of strings and quintessence in different values of parameters with $\omega=-\frac{1}{3}$.}
	\end{figure}

	For a particle of energy $E$ and angular momentum $L$ residing in a
	system with temperature $T$ and angular velocity $\omega_L$, the Maxwell–
	Boltzmann distribution is\cite{Wang:2015zpa,Tao:2015dhe}
	\begin{equation}
		P \propto \exp{(-\frac{E\omega_L L}{T})},
	\end{equation}
	and the probability of a particle tunneling from inside to outside of the horizon is given\cite{Hawking:1975vcx,Unruh:1976db}
	\begin{equation}
		P_{emit}\propto \exp (-\frac{2ImI_{+}-2ImI_{-}}{\hbar}).
	\end{equation}
	Effective Hawking temperature can be read off
	from the Boltzmann factor in $P_{emit}$
	\begin{equation}\label{9}
		T=\frac{\hbar E}{2(ImI_{+}-ImI_{-})}.
	\end{equation}
	For the RN-AdS black hole with cloud of strings and quintessence, we have
	\begin{equation}
		ImI_{\pm}=\pm \frac{\pi E}{f'(r_{+})},
	\end{equation}
	and
	\begin{equation}
		T_0=\frac{\hbar f'(r_{+})}{4 \pi}.
	\end{equation}
	So, from the Beckenstein-Hawking entropy theory\cite{Bekenstein:1973ur}, the entropy can be written as
	\begin{equation}\label{11}
		S_0=\pi r_{+}^2.
	\end{equation}

	From Fig.1, the trend of the Hawking temperature has changed with the raise of the values of different parameters. In Fig.1(a), a local maximum of the temperature exists. However, with $M$ increasing, the local maximum drops and then vanishes, which means the Hawking temperature tends to increase steadily with $M$ raises. And the temperature drops with the increase of $M$, which means that the evaporation of the black hole slows down. In Fig.1(b) and Fig.1(c), the temperature drops with the increase of $a$ and $\alpha$, which is the same as Fig.1(a). In Fig.1(d), as three parameters increase together, we can obtain that the increase of $M$, $a$, and $\alpha$ is to slow down the evaporation of the black hole and make the Hawking temperature rise more steadily.

	\section{Thermodynamics of the RN-AdS black hole with cloud of strings and quintessence in stationary frame in rainbow gravity}\label{444}	
	There are two functions called rainbow function $F(x)$ and $G(x)$ with the following properties
	\begin{equation}
		\lim\limits_{x\to0}F(x)=1,\qquad \lim\limits_{x\to0}G(x)=1.
	\end{equation}
	Specifically, the modified energy-momentum dispersion relation of particle energy $E$ and momentum $p$ can take the following form
	\begin{equation}\label{5}
		E^2 F(E/E_p)^2-p^2 G(E/E_p)^2=m^2,
	\end{equation}
	where $E_{p}$ is the Planck energy, 
	and we set
	\begin{equation}
		\frac{E}{E_{p}}\equiv x.
	\end{equation}
	One of the most popular choices for the functions $F(x)$ and $G(x)$ has been proposed by Amelino-Camelia\cite{Amelino-Camelia:1996bln, Amelino-Camelia:2008aez}, which gives
	\begin{equation}\label{6}
		\begin{aligned}
			F(x)=1,
			G(x)=\sqrt{1-\eta x^n}.
		\end{aligned}
	\end{equation}
	First, we can use the Heisenberg uncertainty principle to estimate the momentum $p$ of an emitted particle\cite{Bekenstein:1973ur, Adler:2001vs}
	\begin{equation}\label{7}
		p \sim \delta p \sim \frac{\hbar}{\delta x} \sim \frac{\hbar}{r_{+}} \sim \frac{1}{r_{+}}.
	\end{equation}
	Combine equation(\ref{5}) with equation(\ref{6}) and equation(\ref{7}), then we have
	\begin{equation}
		\begin{aligned}
			&n=2,E_{n=2}=\frac{\sqrt{m^2 r_{+}^2+1}}{\sqrt{\eta +r_{+}^2}},\\
			&n=4,E_{n=4}=\frac{\sqrt{\frac{\sqrt{4 \eta +4 \eta  m^2 r_+^2+r_+^4}}{\eta }-\frac{r_+^2}{\eta }}}{\sqrt{2}}.
		\end{aligned}
	\end{equation}

	The metric of the static RN-AdS black hole with cloud of strings and quintessence can be written as\cite{Gangopadhyay:2016rpl, Haldar:2019fcz, Lobo:2019jdz}
	\begin{equation}
		ds^2 = -\frac{f(r)}{F(x)^2}dt^2 +\frac{1}{f(r) G(x)^2} dr^2 + \frac{r^2}{G(x)^2} (d\theta^2 + \sin ^2 \theta d\phi^2).
	\end{equation}
	Using the Hamilton-Jacobi method, we can get
	\begin{equation}
		Pr_{\pm}=\pm \left[f(r) G(x)\right]^{-1} \left\{E^2-f(r)\left[G(x) \frac{\lambda}{r^2}+m^2\right]\right\}^{\frac{1}{2}}.
	\end{equation}
	Take this equation into equation(\ref{8}) and we can obtain the Hawking temperature
	\begin{equation}
		T=\frac{f'(r_{+}) G(x)}{4 \pi}=T_0 G(x),
	\end{equation}
	and the entropy can be obtained via equation(\ref{12}) and equation(\ref{11}). For $n=2$, we can get a strict solution of the entropy
	\begin{equation}
		S=\frac{\pi  \left(r_{+} \sqrt{\eta +r_{+}^2}+\eta  \log \left(\sqrt{\eta +r_{+}^2}+r_{+}\right)\right)}{\sqrt{1-\eta  m^2}}.
	\end{equation}

%
	

	\begin{figure}[htbp]
		\centering
		\subfigure[$n=2$]{
			\includegraphics[scale=0.61]{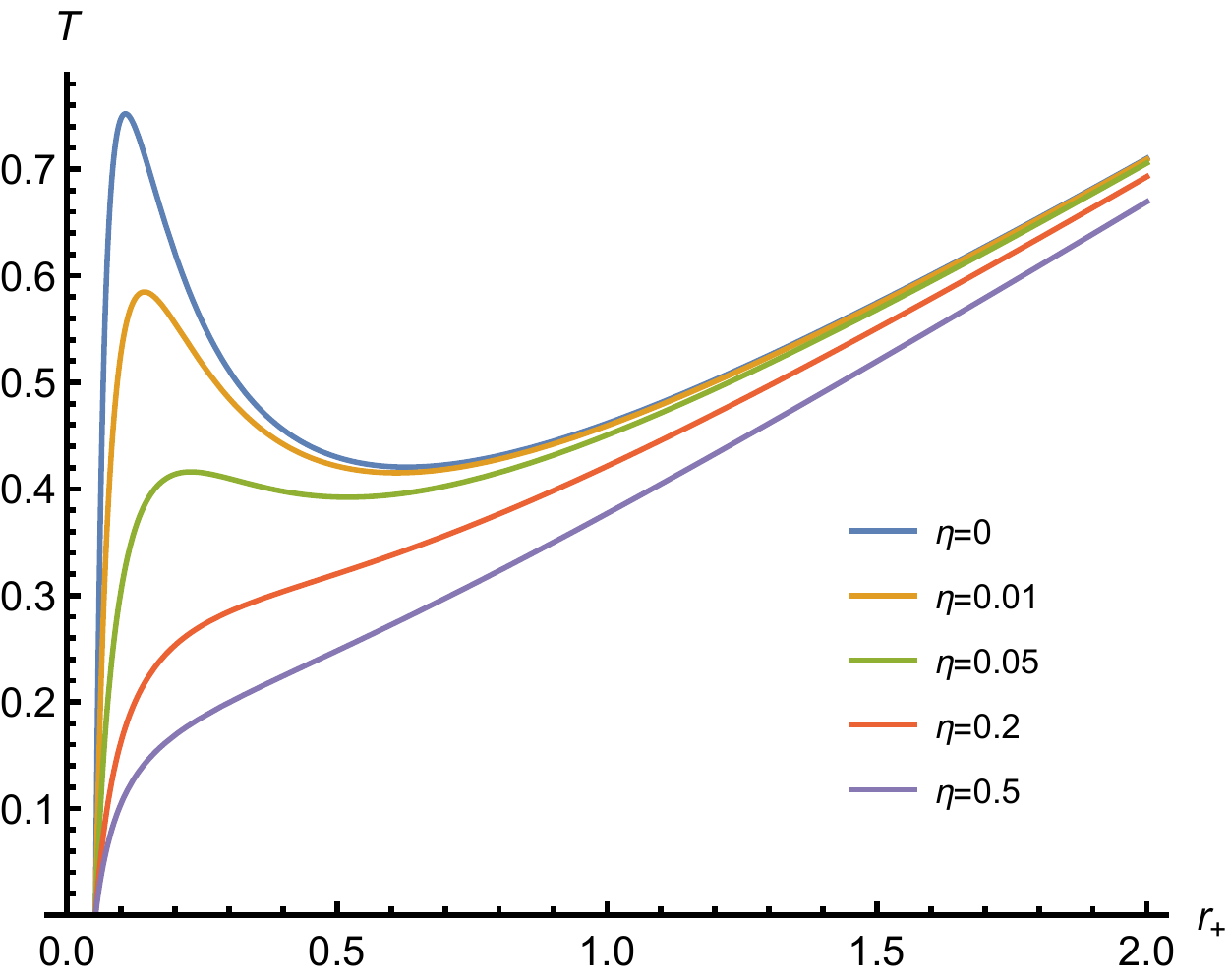}}
		\quad
		\subfigure[$n=4$]{
			\includegraphics[scale=0.61]{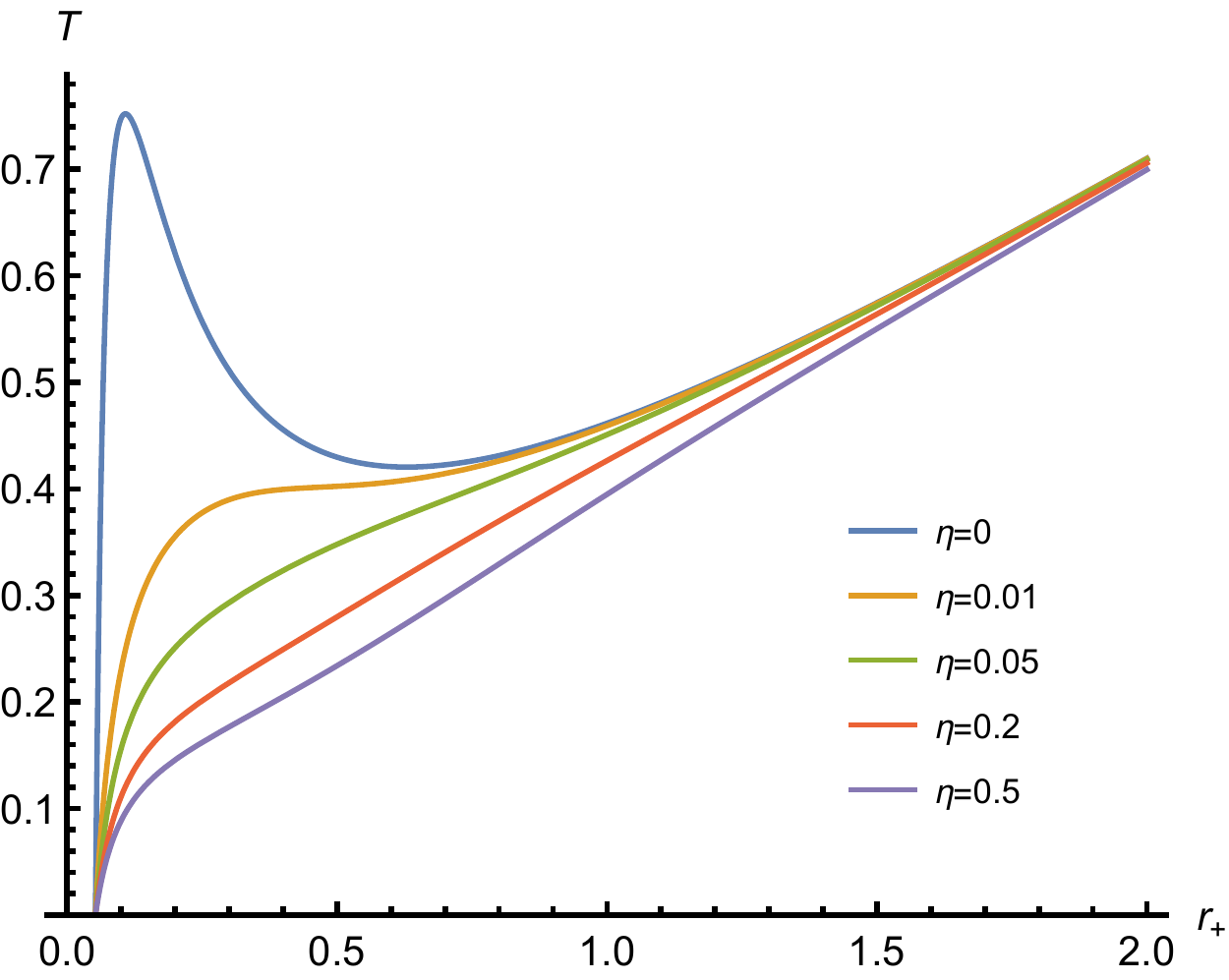}}
		\caption{Plots of the Hawking temperature $T(r)$ of a ST RN-AdS black hole with cloud of strings and quintessence with various values of $\eta$. Here we set $\omega=-\frac{1}{3}$, $M=0.05$, $a=0.025$ and $\alpha=0.025$.}
	\end{figure}
	
	\begin{figure}[htbp]
		\centering
		\subfigure[$n=2$]{
			\includegraphics[scale=0.61]{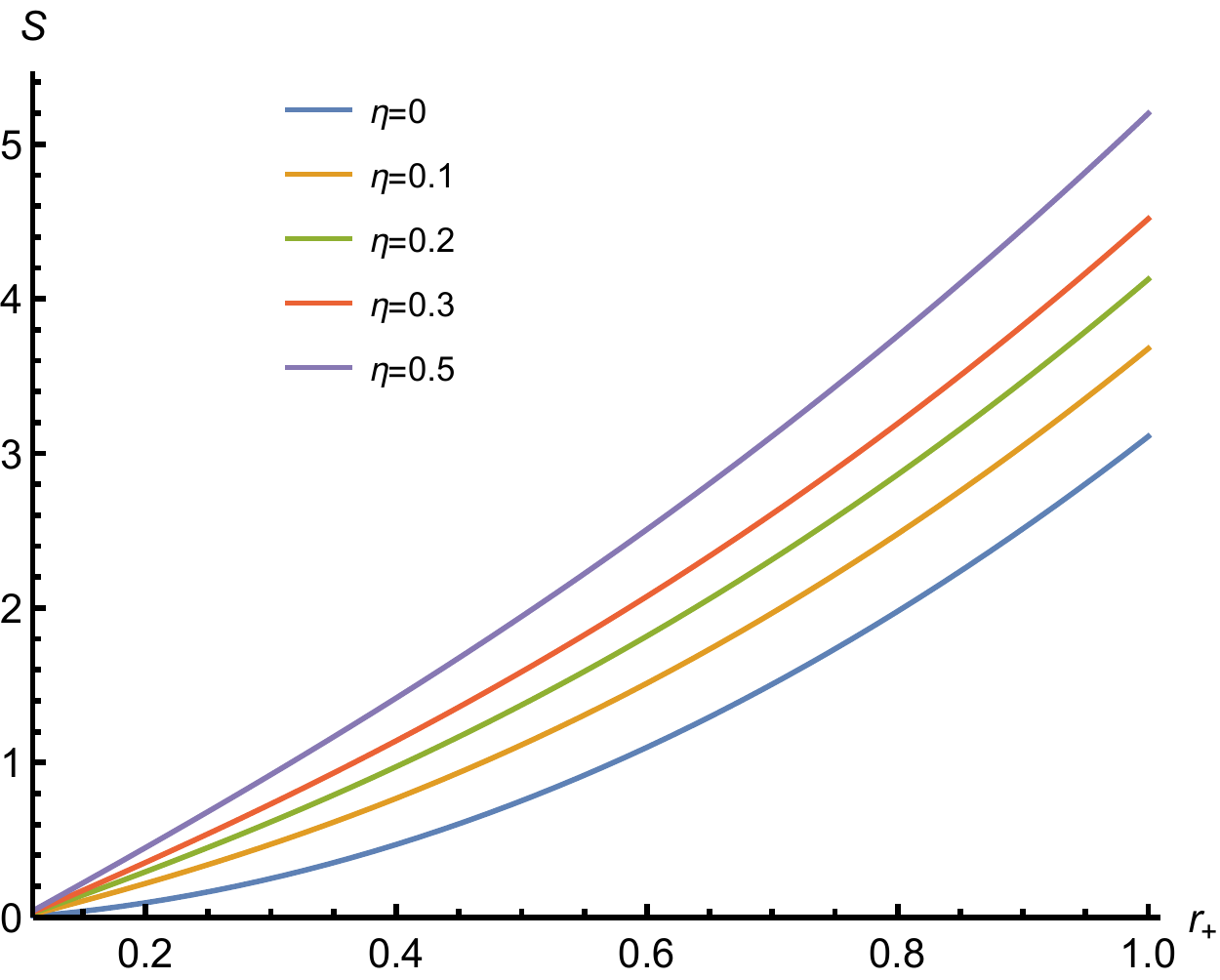}}
		\quad
		\subfigure[$n=4$]{
			\includegraphics[scale=0.61]{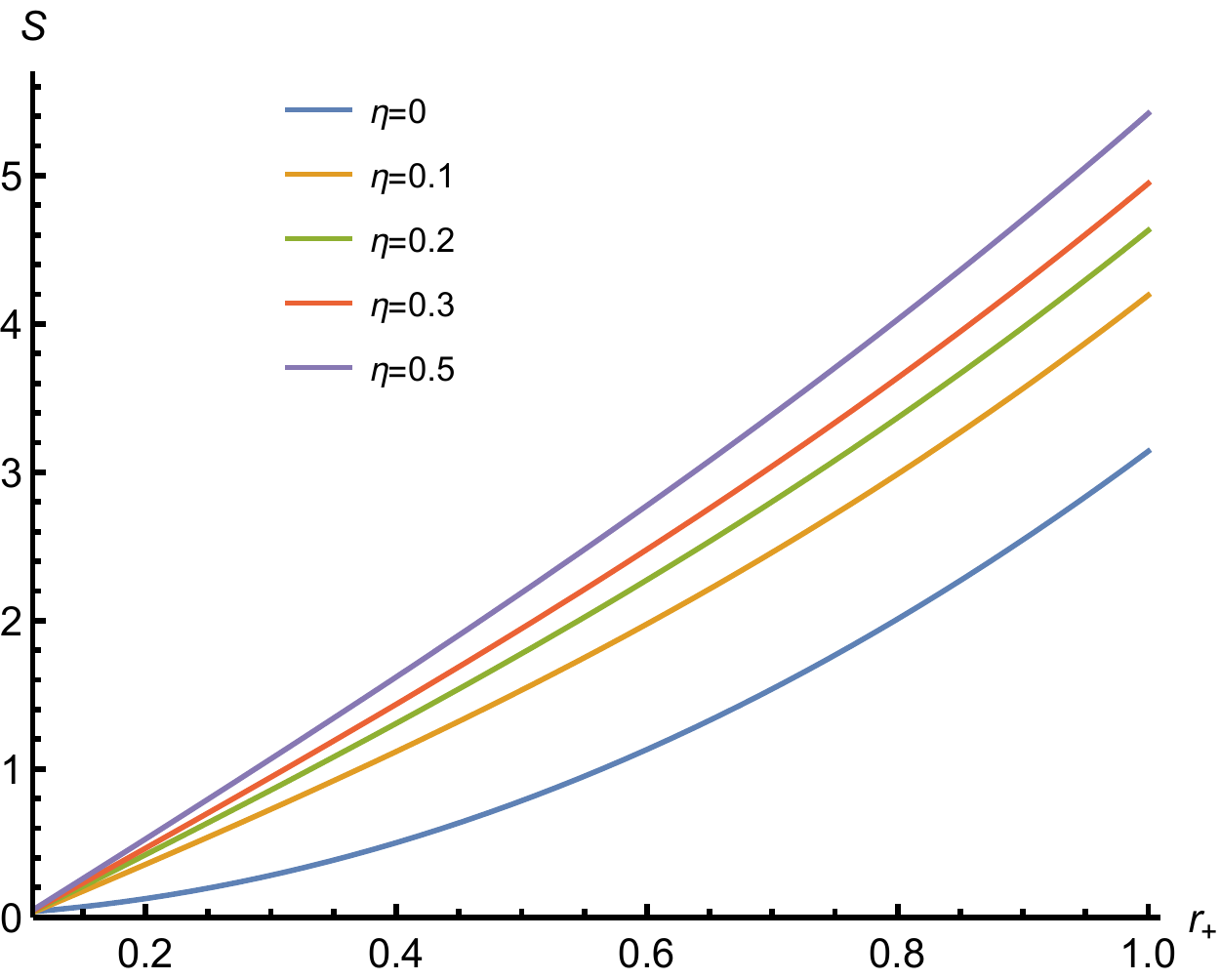}}
		\caption{Plots of the entropy $S(r)$ of a ST RN-AdS black hole with cloud of strings and quintessence with various values of $\eta$. Here we set $\omega=-\frac{1}{3}$, $M=0.05$, $a=0.025$ and $\alpha=0.025$.}
	\end{figure}
	
	To numerically investigate the Hawking temperature $T (r_+)$ and the entropy $S (r_+)$ in ST frame, we focus on the Amelino-Camelia dispersion relation with $n = 2$ and $\eta > 0$.
	For $T (r_+)$ in Fig.2 with $n=2$, the temperature decreases with the increase of $\eta$, which means that the effects of rainbow gravity in SF frame are to slow down the evaporation of the black hole, and the result is similar with increasing the origin parameters.
	For $S (r_+)$ in Fig.3 with $n=2$, the entropy increases with the increase of $\eta$, which means the black hole tends to store more information.
	Meanwhile, for $n=4$, the effects of rainbow gravity on the Hawking temperature and the entropy are like $n=2$.
	From Fig.2 and Fig.3, we can find that the effects of rainbow gravity with $\eta$ increasing in SF frame are to slow down the evaporation of the black hole and make the black hole tends to store more information.

	\section{Thermodynamics of the RN-AdS black hole with cloud of strings and quintessence in free-fall frame in rainbow gravity}\label{666}

	Because of the spherically symmetric metric, the angular momentum of particle is conserved, we can use the fact that $p_t$ and $p_{\phi}$ are conserved along geodesics, since the metric does not depend explicitly on $t$ and $\phi$. The 4-velocity is given by
	\begin{equation}
		\begin{aligned}
			&u_{t}=g_{tt}u^{t}=-\frac{E}{m},\\
			&u_{\phi}=g_{\phi \phi}u^{\phi}=\frac{L}{m},
		\end{aligned}
	\end{equation}
	where $E$, $L$ and $m$ is the energy, angular momentum and mass of the absorbed particle. Solve the equations and we have
	\begin{equation}
		\begin{aligned}
			&u^{t}=\dot{t}=\frac{1}{f(r)} \left(\frac{E}{m}\right),\\
			&u^{\phi}=\dot{\phi}=\frac{1}{r^2 \sin ^2 \theta} \left(\frac{L}{m}\right).
		\end{aligned}
	\end{equation}
	For $u^{\mu}$, we have $g^{\mu \nu} u_{\mu} u_{\nu}=-1$, which becomes
	\begin{equation}
		g^{tt} u_{t}^2 + g^{rr} u_{r}^2 + g^{\phi \phi} u_{\phi}^2 = -1.
	\end{equation}
	Solve it and we can get
	\begin{equation}
		u^{r}=g^{rr} u_r=\sqrt{f(r) \left[-1+\frac{1}{f(r)} \left(\frac{E}{m}\right)^2-\frac{1}{r^2}  \left(\frac{L}{m}\right)^2\right]}.
	\end{equation}
	
	The energy-independent rainbow metric of the RN-AdS black hole with cloud of strings and quintessence is given by\cite{MahdavianYekta:2019dwf}
	\begin{equation}
		ds^2=g_{\mu\nu}dx^{\mu}\otimes dx^{\nu},
	\end{equation}
	which can be rewritten in terms of the energy-independent orthonormal frame fields
	\begin{equation}
		ds^2=\eta^{ab} e_a \otimes e_b,
	\end{equation}
	where $a,b=(0,i)$, and $i$ is the spatial index. So for the MDR, the energy-dependent rainbow counterpart for the energy-independent metric can be obtained using equivalence principle\cite{Magueijo:2002xx}
	\begin{align}
		\begin{aligned}
			d\widetilde{s}^2&=\widetilde{g}_{\mu\nu}dx^{\mu}\otimes dx^{\nu}=\eta^{ab} \widetilde{e}_a \otimes \widetilde{e}_b \\
			&=\left(\frac{1}{G^2(x)}-\frac{1}{F^2(x)}\right)e_{0}\otimes e_{0}+\frac{ds^2}{G^2(x)},
		\end{aligned}
	\end{align}
	where $\frac{E}{E_{p}}\equiv x$ and the energy-dependent orthonormal frame field can be written as
	\begin{align}
		\widetilde{e}_0=\frac{e_0}{F(x)},\widetilde{e}_i=\frac{e_i}{G(x)}.
	\end{align}
	So, the free-fall rainbow metric of the RN-AdS black hole with cloud of strings and quintessence can be obtained
	\begin{align}
		d\widetilde{s}^2=
		\left(\frac{1}{G^2(x)}-\frac{1}{F^2(x)}\right)e_{0}\otimes e_{0}+\frac{-f(r) dt^2 +\frac{1}{f(r)} dr^2 + r^2 (d\theta^2 + \sin ^2 \theta d\phi^2)}{G^2(x)},
	\end{align}
	where $e_0=u_{\mu} dx^{\mu}$. From the 4-velocity we have
	\begin{equation}
		\begin{aligned}
			&g^{tt}=\frac{\frac{E^2 \left[G(x)^2-1\right]}{m^2}-f(r) G(x)^2}{f(r)^2},\\
			&g^{rr}=f(r)+\frac{E^2 \left[G(x)^2-1\right]}{m^2},\\
			&g^{tr}=g^{rt}=\frac{E \left[G(x)^2-1\right] \sqrt{\frac{E^2-f(r) m^2}{f^2 m^2}}}{m},\\
			&g^{\theta \theta}=\frac{G(x)^2}{r^2},\\
			&g^{\phi \phi}=\frac{G(x)^2}{r^2}.
		\end{aligned}
	\end{equation}
	So, with equation(\ref{3}) we can get
	\begin{equation}
		Pr_{\pm}=\frac{B \pm C}{A},
	\end{equation}
	where
	\begin{equation}
		\begin{aligned}
			&A=2 (f+\frac{E^2 \left[G(x)^2-1\right]}{m^2}),\\
			&B=\frac{2 E^2 \left(G(x)^2-1\right) \sqrt{\frac{E^2-f(r) m^2}{f(r)^2 m^2}}}{m},\\
			&C=\frac{2 \sqrt{-f(r) G(x)^2 \lambda  m^2-f(r) m^4 r^2-E^2 G(x)^4 \lambda +E^2 G(x)^2 \lambda +E^2 m^2 r^2}}{m r}.
		\end{aligned}
	\end{equation}

	Take this equation into equation(\ref{8}) andequation(\ref{9}) we can calculate the Hawking temperature
	\begin{equation}\label{13}
		\begin{aligned}
			\widetilde{T}&=\frac{T_0}{1 + \epsilon  (\frac{E^2 f''(r_{+})}{m^2 f'(r_{+})^2}-\frac{1}{2})}\\
			&=\frac{1}{4 \pi } \left(\frac{2 r_{+}}{l^2}+\frac{2 M}{r_{+}^2}-\frac{2 Q^2}{r_{+}^3}+\frac{\alpha  (3 \omega +1)}{r_{+}^{3 \omega +2}}\right) \\
			&\times \left(1-\epsilon  \left(\frac{E^2 (-\frac{2}{l^2}-\frac{4 M}{r_{+}^3}+\frac{6 Q^2}{r_{+}^4}-\alpha  (-3 \omega -2) (-3 \omega -1) r_{+}^{-3 \omega -3})}{m^2 (-\frac{2 r_{+}}{l^2}+\frac{2 M}{r_{+}^2}-\frac{2 Q^2}{r_{+}^3}-\alpha  (-3 \omega -1) r_{+}^{-3 \omega -2})^2}-\frac{1}{2} \right) \right),
		\end{aligned}
	\end{equation}
	where $\epsilon=\eta (\frac{E}{E_p})^n$.
	Equation(\ref{12}), (\ref{11}) and (\ref{13}) give out the entropy 
	\begin{equation}
	\frac{d\widetilde{S}}{dr}=2 \pi r \left( 1 + \epsilon  \left(\frac{E^2 f''(r_{+})}{m^2 f'(r_{+})^2}-\frac{1}{2}\right) \right).
	\end{equation}

	\begin{figure}[htbp]
		\centering
		\subfigure[$n=2$, $M=10$]{
			\includegraphics[scale=0.55]{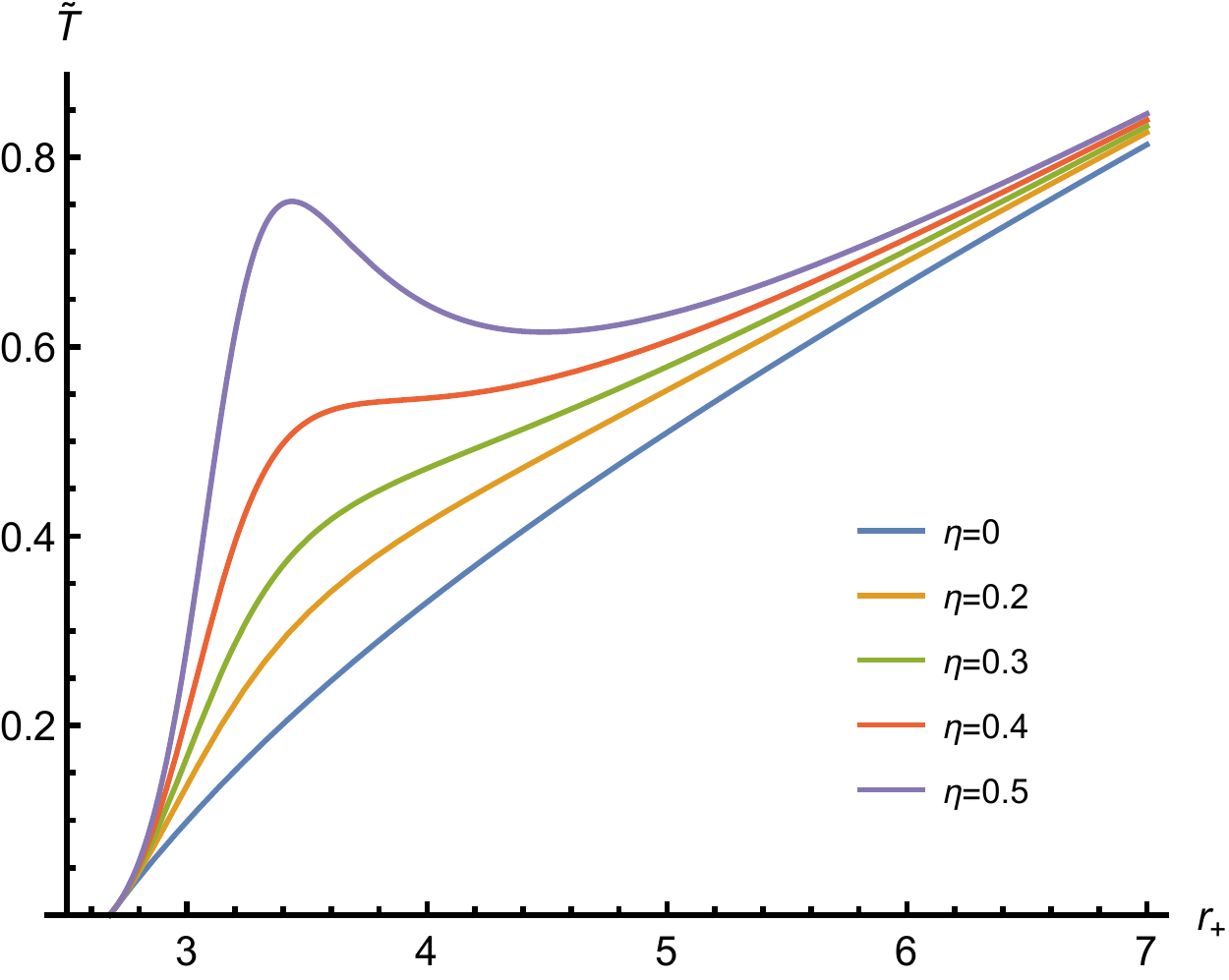}}
		\quad
		\subfigure[$n=4$, $M=2.4$]{
			\includegraphics[scale=0.55]{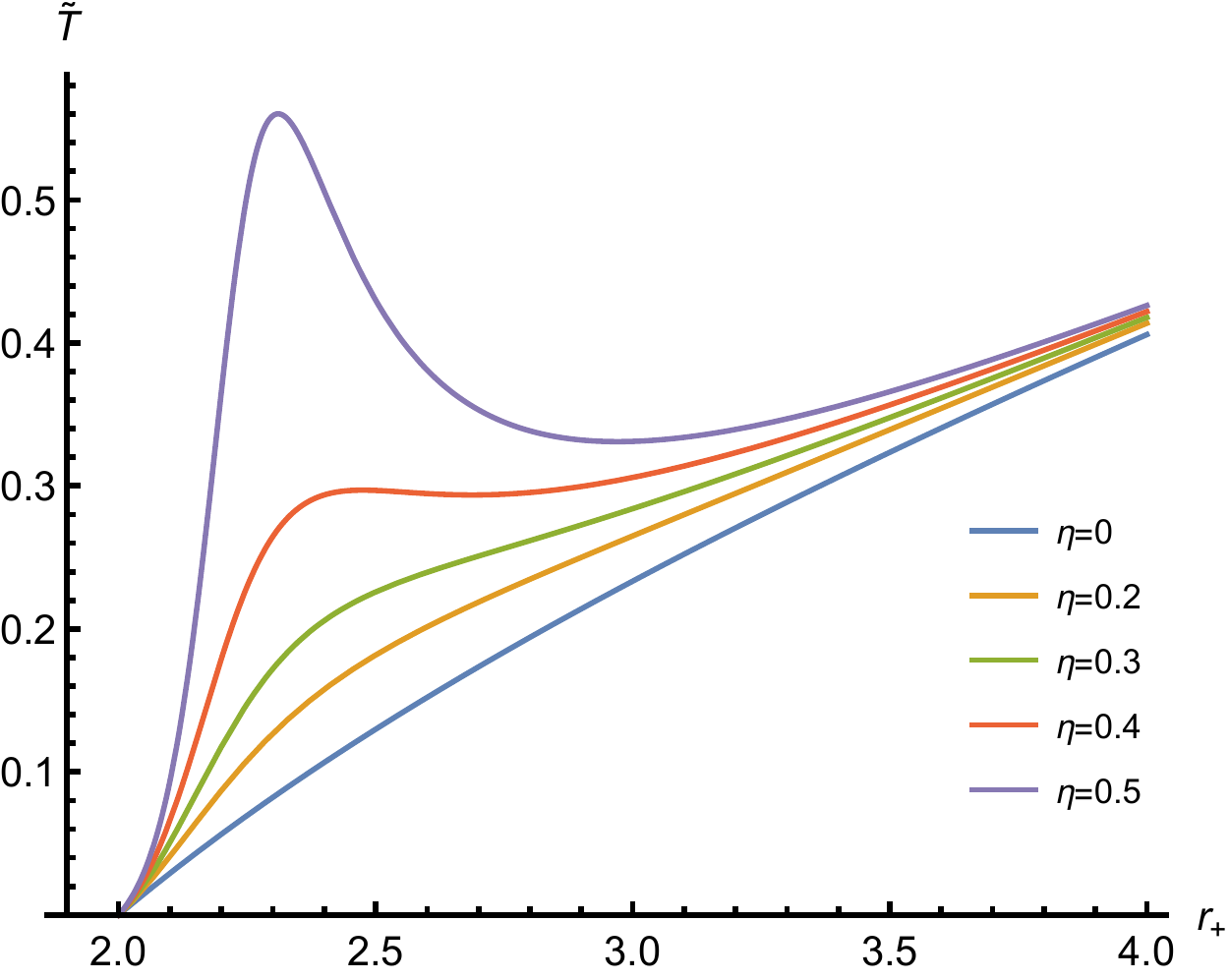}}
		\caption{Plots of the Hawking temperature $T(r)$ of a FF RN-AdS black hole with cloud of strings and quintessence with various values of $\eta$. Here we set $\omega=-1$, $a=3$ and $\alpha=0.6$.}
	\end{figure}
	
	\begin{figure}[htbp]
		\centering
		\subfigure[$n=2$, $M=10$]{
			\includegraphics[scale=0.55]{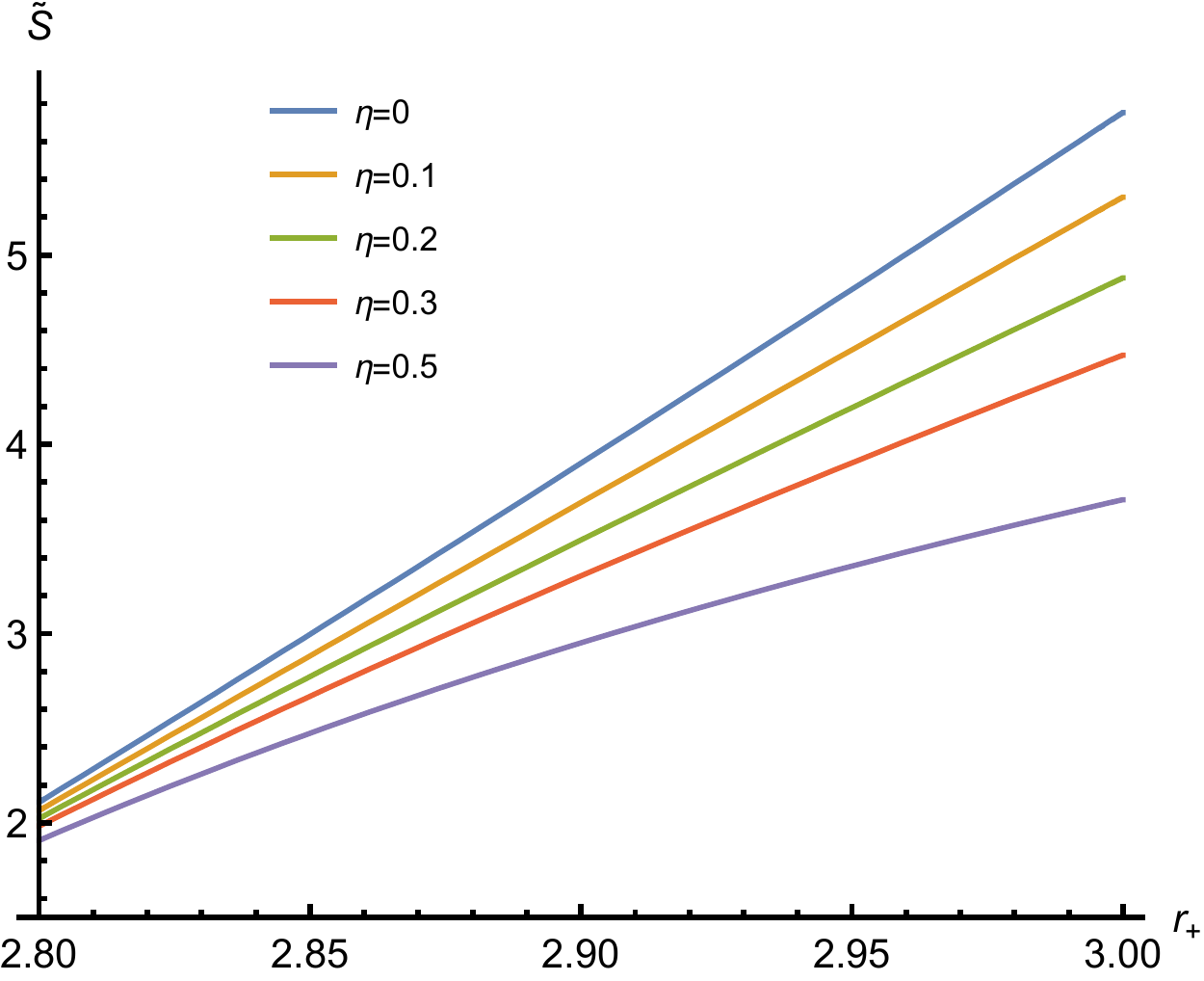}
			\includegraphics[scale=0.55]{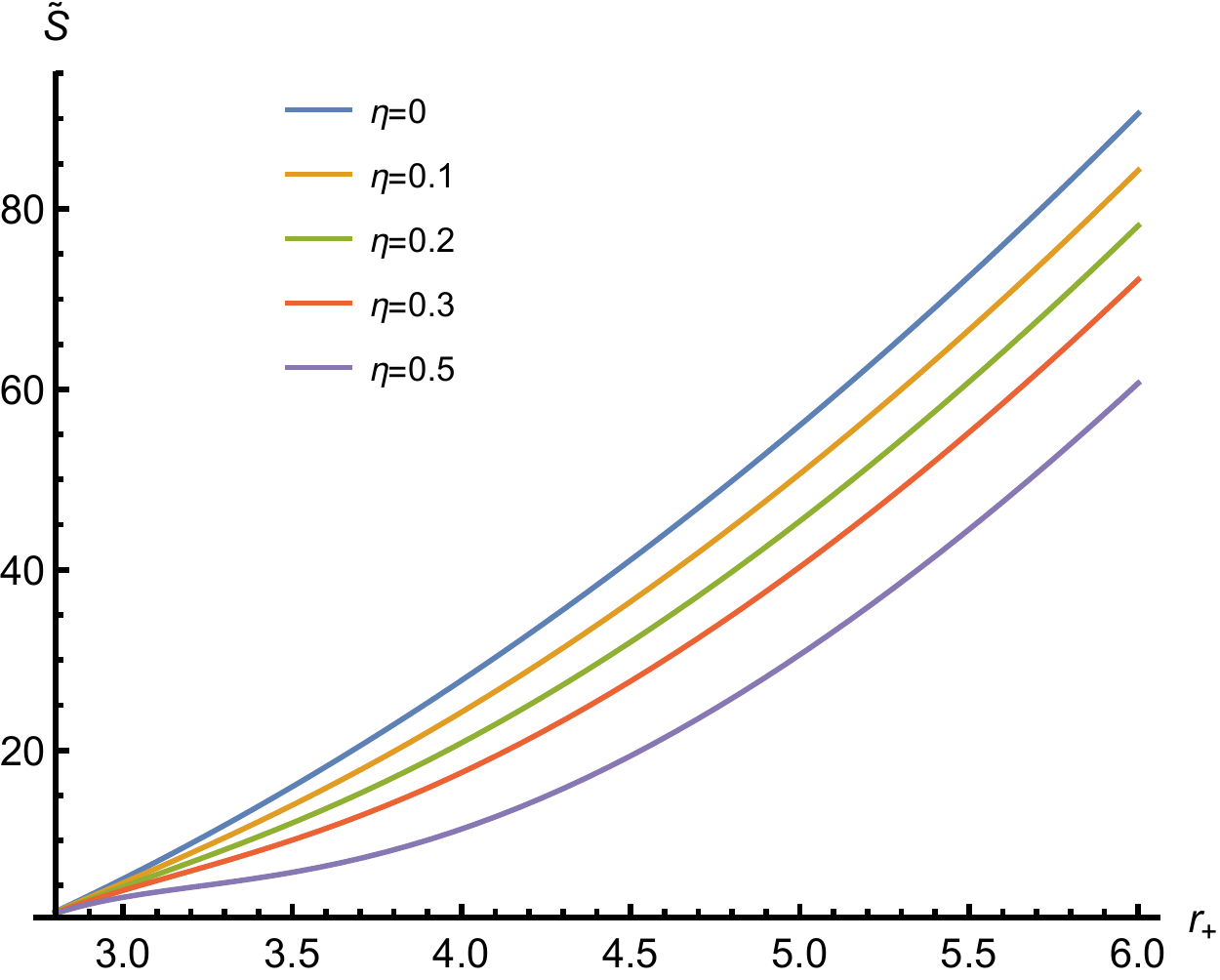}}
		\quad
		\subfigure[$n=4$, $M=2.4$]{
			\includegraphics[scale=0.55]{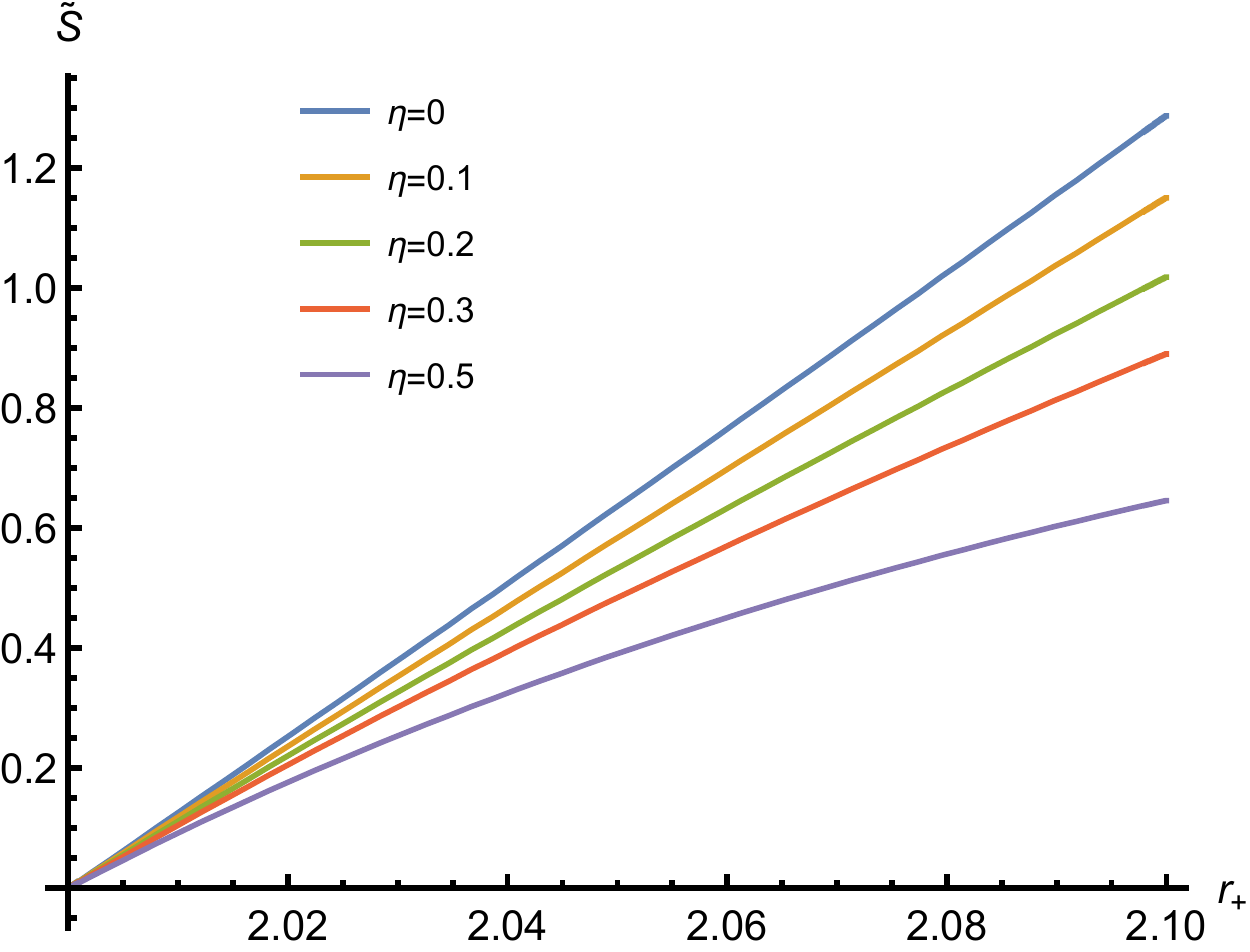}
			\includegraphics[scale=0.55]{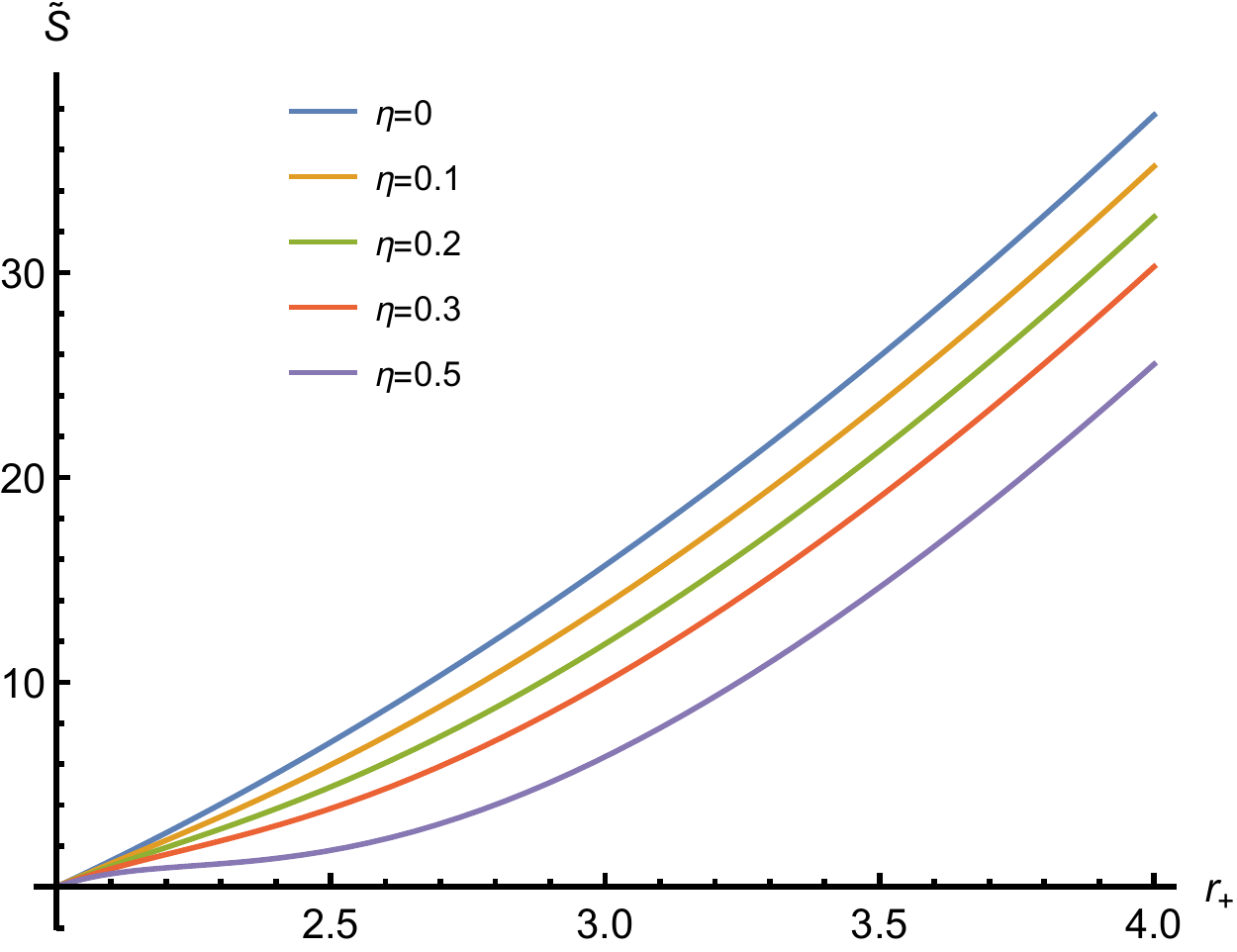}}
		\caption{Plots of the entropy $S(r)$ of a FF RN-AdS black hole with cloud of strings and quintessence with various values of $\eta$. Here we set $\omega=-1$, $a=3$ and $\alpha=0.6$.}
	\end{figure}
	
	Through the similar way, we obtain the modified Hawking temperature $\widetilde{T} (r_+)$ and black hole entropy $\widetilde{S} (r_+)$ in FF frame by analytical and numerical method with $n = 2$ and $\eta > 0$.
	For $\widetilde{T} (r_+)$ in Fig.4 with $n=2$, the Hawking temperature decrease with the increase of $\eta$, meaning that the effects of rainbow gravity in FF frame are to speed up the evaporation of the black hole.
	For $\widetilde{S} (r_+)$ in Fig.5 with $n=4$, the entropy decreases with the increase of $\eta$, meaning that the black hole tends to store less information under the rainbow gravity in FF frame with higher $\eta$.
	Meanwhile, for $n=4$, the effects of rainbow gravity on the Hawking temperature and the entropy is like $n=2$.
	From Fig.4 and Fig.5, we can find that the effects of rainbow gravity with $\eta$ increasing are to speed up the evaporation of the black hole and make the black hole tends to store less information, which seems to be the opposite of the situation in ST frame.

	\section{Discussion and conclution}\label{777}
	In this paper, we consider the RN-AdS black hole with cloud of strings and quintessence in ST frame and FF frame in rainbow gravity and analyze the difference of these two frames on the Hawking temperature and the entropy.
	We find that, for ST frame, the effects of rainbow gravity are to make the Hawking temperature rises more steadily and slow down the evaporation of the black hole. Meanwhile, the entropy increases, making the black hole tend to store more information.
	However, for FF frame, the effects of rainbow gravity are to make the Hawking temperature rises less steadily and speed up the evaporation of the black hole. Meanwhile, the entropy decreases, making the black hole tend to store less information.
	Moreover, as cloud of strings and quintessence bring two parameters to the metric of the black hole, the effect of these two parameters increasing is similar to the effect of the balck hole's mass increasing, which means that cloud of strings and quintessence tend to make the rainbow effects more obvious.
	
	\begin{table}[htbp]\label{50}
		\centering
		\caption{Results and implications for the Hawking temperature and the entropy of the ST and FF rainbow RN-AdS black hole with cloud of strings and quintessence.}
		
		\begin{tabular}{p{2.5cm}|p{7cm}|p{7cm}}
			\hline
			&FF rainbow black hole&ST rainbow black hole\\
			\hline
			Temperature
			&The rainbow effects increase the temperature and make it rise more steadily, which means rainbow speeds up the evaporation of the black hole.
			&The rainbow effects decrease the temperature and make it rise less steadily, which means rainbow slows down the evaporation of the black hole.\\
			\hline
			Entropy
			&The rainbow effects decrease the entropy, which means the black hole tends to store less information.
			&The rainbow effects increase the entropy, which means the black hole tends to store more information.\\
			\hline
		\end{tabular}
	\end{table}

	We compare the RN-AdS black hole with cloud of strings and quintessence with the Schwarzschild black hole. The Schwarzschild black hole under rainbow gravity has been discussed in the ST and FF cases\cite{Mu:2015qna, Tao:2016baz}. For the Schwarzschild black hole, the rainbow gravity has similar effects in both ST frame and FF frame. It tends to decrease the Hawking temperature and increase the entropy, which means that the evaporation of the black hole will be slowed down, and the black hole tends to store more information.
	However, our results show that the effects of the FF rainbow RN-AdS black hole with cloud of strings and quintessence behave quite opposite to the ST rainbow case. It seems that the effects rainbow gravity has are quite model-dependent.
	It would be interesting to study the effects of the rainbow RN-AdS black hole with cloud of strings and quintessence in cavity in the future, which might be useful to explore more remarkable effects of the rainbow gravity.

	\section*{Acknowledgement}
	We are grateful to Haitang Yang, Peng Wang and Yuhang Lu for useful discussions. This work is supported in part by National Science Foundation of China (Grant No. 11747171), Natural Science Foundation of Chengdu University of TCM (Grants No. ZRYY1729 and No. ZRYY1921), Discipline Talent Promotion Program of /Xinglin Scholars (Grant No. QNXZ2018050), and the key fund project for Education Department of Sichuan (Grant No. 18ZA0173). The authors contributed equally to this work.

\end{document}